\documentclass[fleqn,12pt,twoside]{article}
\usepackage{espcrc1}
\usepackage{epsf}
\usepackage{graphicx}
\usepackage{epsfig}

\newcommand{\beq}[1]{
\begin{equation}\label{#1}}
\newcommand{\eeq}{\end{equation}}
\newcommand{\bea}[1]{
\begin{eqnarray}\label{#1}}
\newcommand{\eea}{\end{eqnarray}}

\newcommand{\AmS}{{\protect\the\textfont2
  A\kern-.1667em\lower.5ex\hbox{M}\kern-.125emS}}

\title{
\hfill {\small CPHT RR-053.0706} \\
\vskip.2in
Hard exclusive reactions and hadron structure}

\author{
        B.~Pire\address{{CPhT},
 \'Ecole Polytechnique, 
        91128 Palaiseau, France}\thanks{UMR 7644 du CNRS}         
        L.~Szymanowski\address{Soltan Institute for Nuclear Studies,
        Ho\.{z}a 69, 00-681 Warsaw, Poland and \\ 
        Universit\'e de Li\`ege, 4000 Li\`ege, Belgium}
        }
       
\begin{document}

\maketitle

\begin{abstract}
The generalized Bjorken regime of exclusive reactions opens new ways to explore the hadron structure.
We shortly review the present status of this domain where generalized parton distributions, generalized distribution amplitudes and transition distribution amplitudes describe various aspects of confinement physics.
\end{abstract}

\section{Generalities}

According to a now well established framework \cite{Muller}, the Bjorken limit of exclusive reactions with a hard probe allows factorization of the amplitudes into a perturbatively calculable subprocess at
quark and gluon level on the one hand and hadronic matrix elements of light cone non local operators  expressed through hadron (eventually generalized \cite{GDA}) distribution amplitudes and 
generalized parton distribution (GPD) (resp. a transition distribution amplitude \cite{TDA}) describing the transition from the baryon target to a baryon (resp. another hadron or photon) on the other hand. 

As demonstrated in the presentations at this workshop, experimental data from DESY and JLab are now confirming this framework, and they seem to show that its applicability is quite precocious in terms of $Q^2$. Important in that respect is the analysis of spin asymmetries which are particularly sensitive to the interference between the dVCS and Bethe-Heitler processes \cite{asym}.


\section{Accessing transversity GPDs}
Accessing chiral-odd hadronic matrix elements is notoriously difficult \cite{tGPD1} . 
We recently showed \cite{tGPD2} that the transversity GPD $H_T(x,\xi,t)$ contributes 
to a measurable electroproduction amplitude. We consider the process
\begin{equation}
  \label{proc-gen}
\gamma^{(*)} N \to   \rho_L \, \rho_T \; N' \;.
\end{equation}
For simplicity
of discussion we specifically study the amplitude for the
process
\begin{equation}
 \label{proc-p}
\gamma^{(*)} p \to \rho_L^0 \, \rho_T^+\; n\;, 
\end{equation}
that is,  virtual or real photoproduction on a
proton $p$, which leads via two-gluon exchange to the production
of  a vector meson $\rho^0$ separated by a large
rapidity gap from another vector meson $\rho^+$ and the scattered neutron $n$.
We consider the kinematical region where the rapidity gap between $\rho^+$ and
$n$ is much smaller than the one between
$\rho^0$ and $\rho^+$, that is the energy of the system ($\rho^+ - n$) is smaller
than the energy of the system ($\rho^0 - \rho^+$) but,  to
justify our approach, still larger
than baryonic resonance masses.
In  such kinematical
circumstances the Born term for this
process is calculable consistently within
the collinear factorization method. The amplitude is represented as an
integral (over  the longitudinal momentum fractions of the quarks)  of
the product of two amplitudes: the first one
 describing the transition $\gamma^{(*)} \to \rho^0_{L}$ via two-gluon exchange and
the second one  describing the pomeron--proton subprocess
${\cal P} p \to  \rho^+ n$ which is
closely related to the electroproduction process $\gamma^*\,N \to \rho \,N'$
where  collinear factorization
theorems allow separating  the long distance dynamics  expressed
through the
GPDs from a perturbatively calculable coefficient function. The hard scale
appearing in the process  is supplied by the
relatively large  momentum transfer
$p^2$ in the two-gluon channel, i.e. by the virtuality of the pomeron.

\begin{figure}[htb]
\centerline{%
\epsfig{file=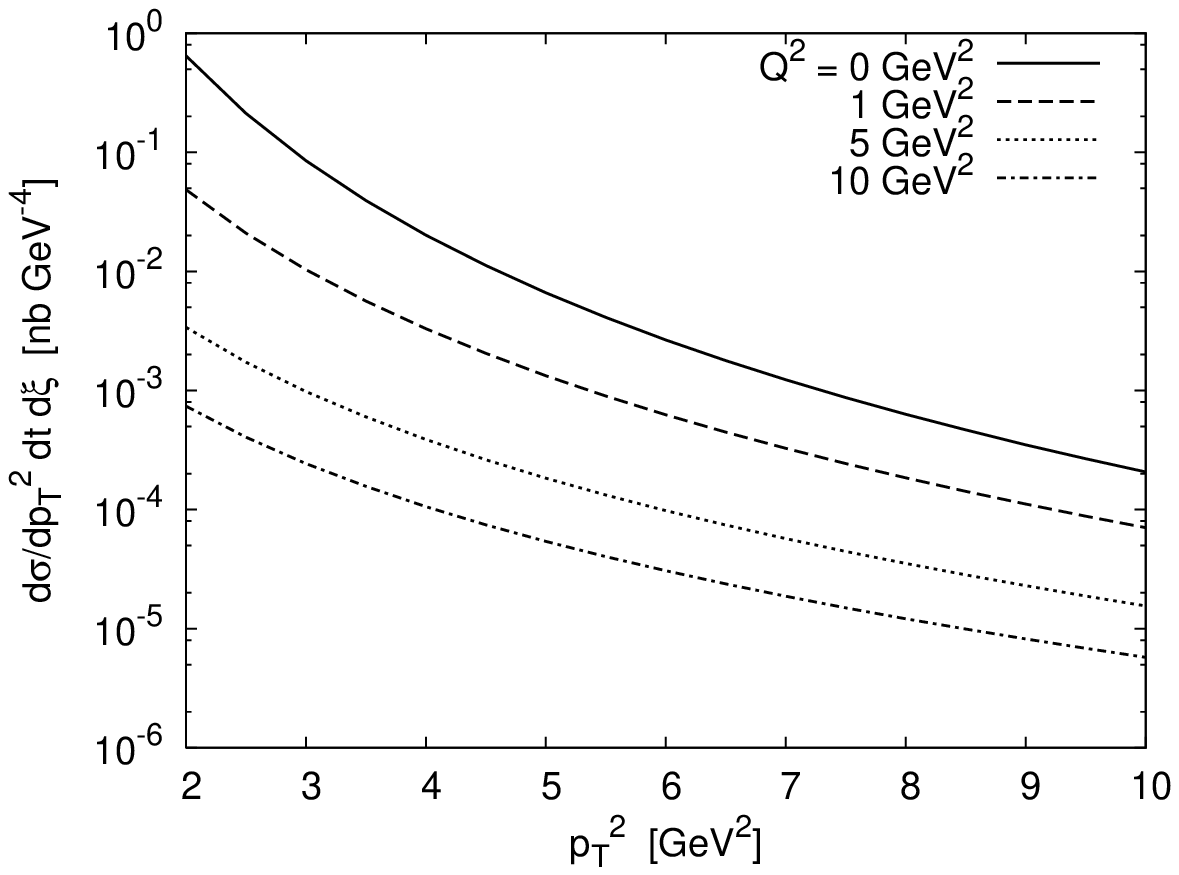,width=0.5\columnwidth}%
\epsfig{file=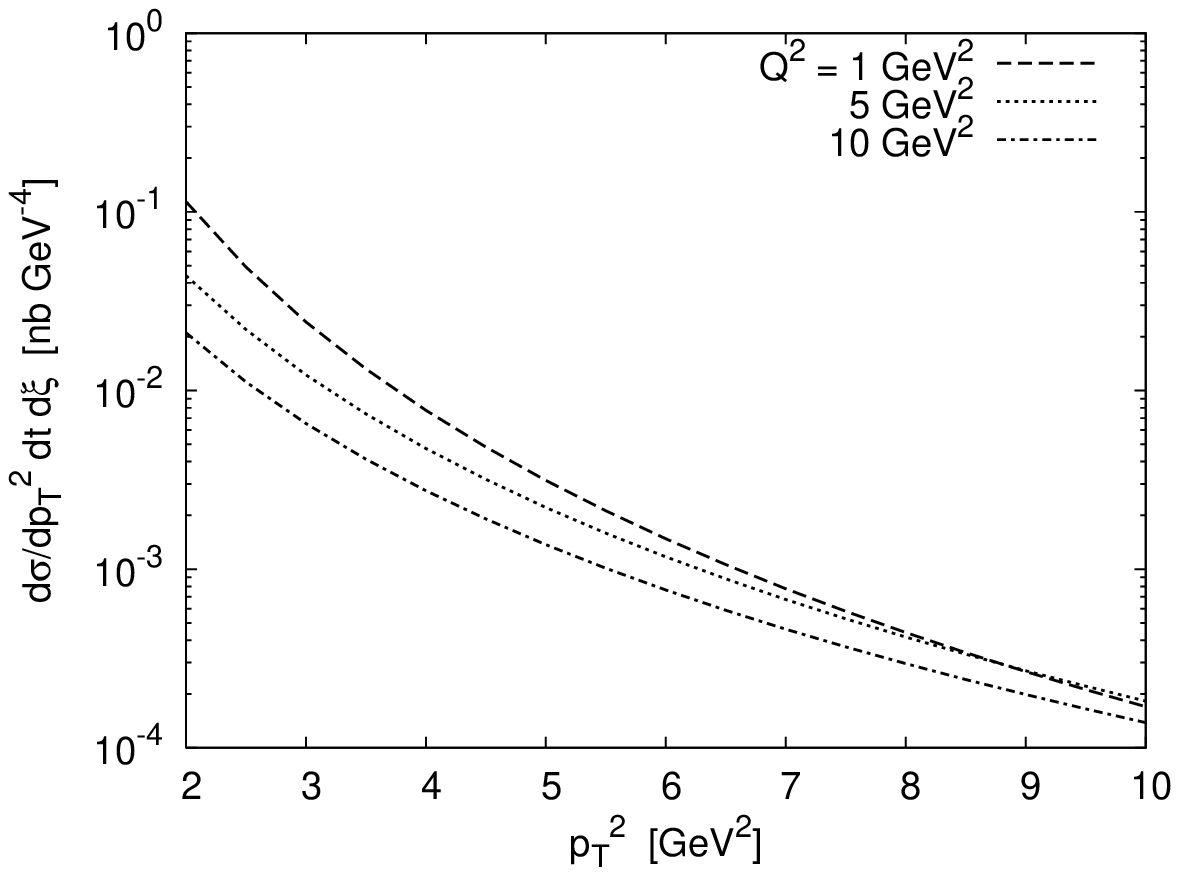,width=0.5\columnwidth}}
\caption{\small 
The differential cross section for $\gamma^*_{L/T}(Q) \,p\to \rho^0_L\,\rho^+_T\,n$ for transverse virtual photon (left) and longitudinal virtual photon (right), plotted as a function of $p_T^2$ for $\xi=0.3$ and $Q^2=$ 0, 1, 5, and 10~GeV$^2$.
 }
\label{Telectro}
\end{figure}

Fig. \ref{Telectro} shows precitions for the  differential cross section for 
$\gamma^*_{L/T}(Q) \,p\to \rho^0_L\,\rho^+_T\,n$ obtained within  a model of
the transversity distribution based on the one meson exchange in the 
t-channel. 
Our estimate is valid  for the high energy limit. If one is willing to 
study the same processes  at lower energy, 
one should include all polarization states of exchanged gluons.

\begin{figure}
\begin{center}
  \epsfxsize 0.4\textwidth
  \epsfbox{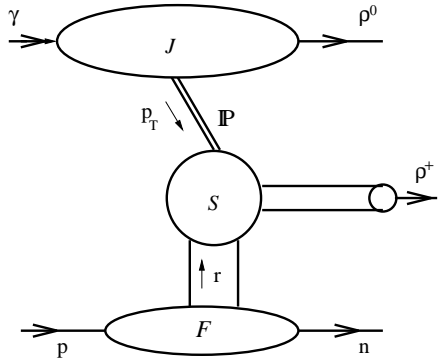}
\end{center}
\caption{\label{fig:process}The photoproduction process to access transversity GPDs.}
\end{figure}
An experimental determination of the transversity GPD $H_{T}$
seems feasible in photo- or electroproduction at high energies 
and we believe that the possible eRHIC 
machine may become the best place where the 
process discussed here could be measured, provided experimental setups allow a large 
angular coverage, ensuring a sufficient detection efficiency and a good control of exclusivity.  The JLab CLAS-12 upgrade probably will have good enough detection efficiency for observing the two rho mesons, but only for relatively low $p_T$ of the order of 1--1.5 GeV. Moreover the smaller energy available prevents the theoretical framework used here from being adequate and  one needs to supplement our studies by adding contributions coming from other polarization states of exchanged gluons and the ones coming from quark exchanges.

%
%

\section{Electroproduction of exotic hadrons}
\begin{figure}
$$\rotatebox{270}{\includegraphics[width=8cm]{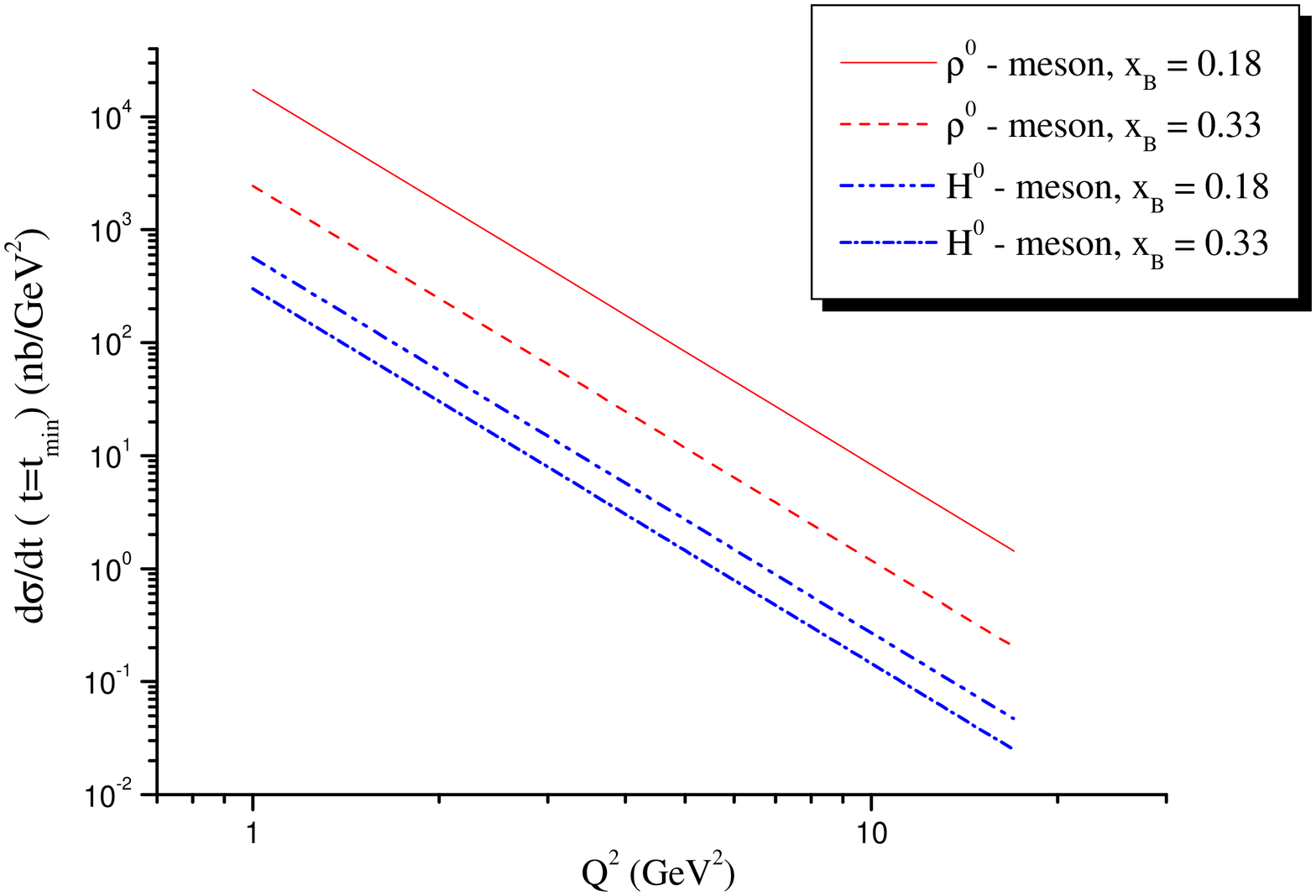}}$$
\caption{
Comparison of the electroproduction 
of exotic and non-exotic vector mesons.}
\label{fighyb}
\end{figure}

Hard production of hadrons outside the quark model has seemed to be a challenge, since 
it was generally believed that these {\em exotic} particles could not have a non-zero leading
twist distribution amplitude (DA). We demonstrated recently \cite{Hybrid} that the non local nature 
of the quark correlators defining a DA was in fact allowing any $J^{PC}$ values for the meson
described by it. Moreover, a relation between the energy-momentum tensor and a moment of this correlator 
allows to estimate the magnitude of the leading twist DA of a $J^{PC}= 1^{-+}$ exotic
vector meson. As seen in Fig.\ref{fighyb} electroproduction cross sections then turn out to be not small in comparison with those of 
usual mesons, and precise data at JLab should thus reveal the properties of these exotic mesons. 

\section{Transition distribution amplitudes}

\begin{figure}[htb]
\centerline{%
\epsfig{file=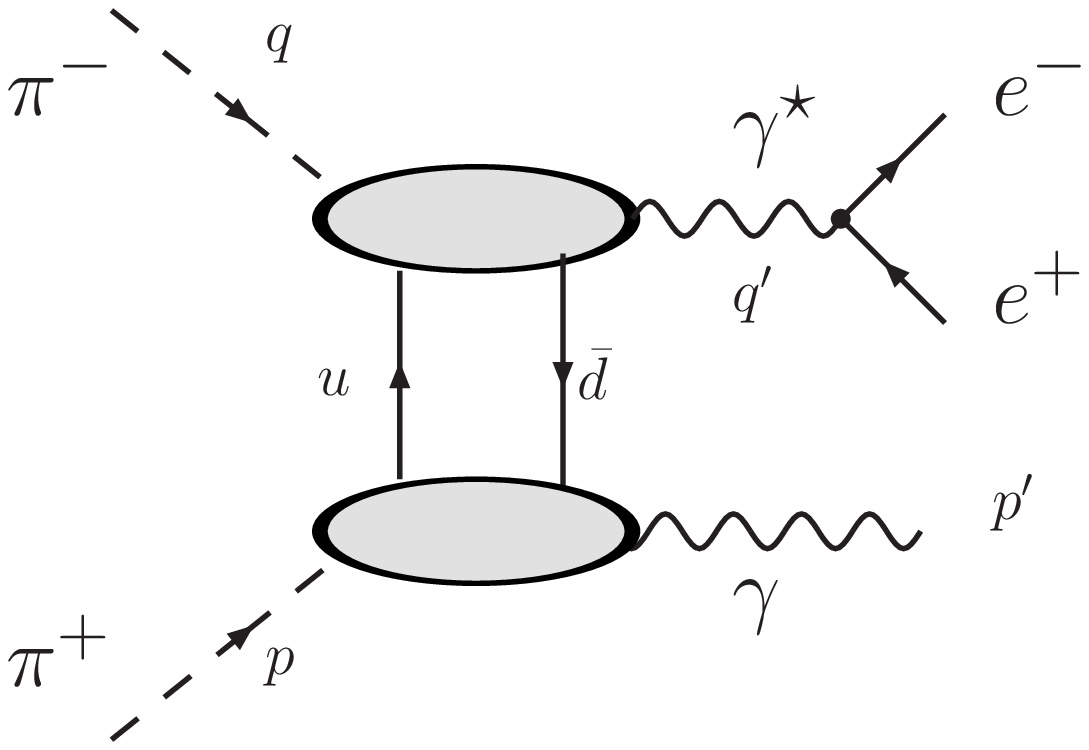,width=0.45\columnwidth}%
\epsfig{file=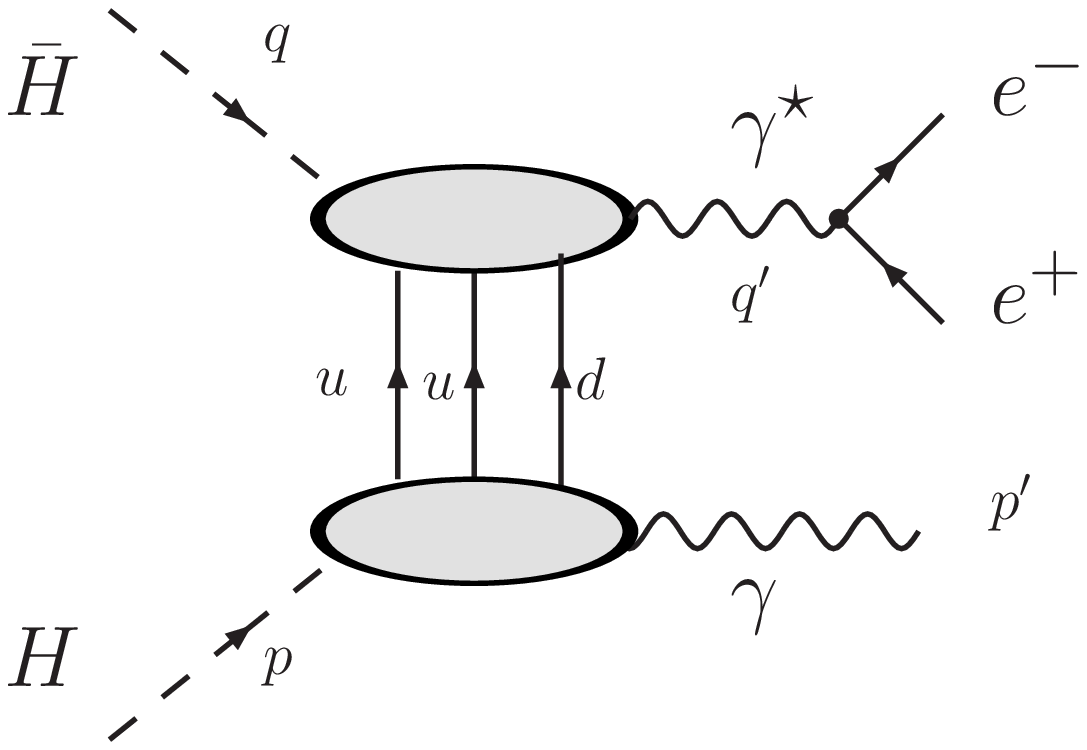,width=0.45\columnwidth}}
\caption{\small 
The factorization of the annihilation process $\bar H \;H\to \gamma^*
\,\gamma$ into a hard subprocess (upper blob) and a transition
distribution amplitude (lower blob) for the meson case  and the baryon case.
 }
\label{TDA1}
\end{figure}

Let us now  consider another class of exclusive processes, such as backward VCS
$e p \to  e\;p \; \gamma $,
where the final photon flies in the direction of the initial proton, or its crossed version
$\bar p p \to  \gamma^*\; \gamma $,
 near the forward direction, which may be studied at GSI-FAIR \cite{PAX}.
 We propose \cite{TDA} that the amplitudes of such processes factorize in a quite similar
  way as for the dVCS reaction, but with a three quark exchange replacing the usual quark 
  antiquark exchange characteristic of the handbag diagrams (see Fig. \ref{TDA1}). For instance, we  write the 
  $\bar p N \to \gamma^*\pi$ amplitude  as 
\begin{equation}
{\cal M} (Q^2,  \xi, t)= \int dx dy \phi(y_i,Q^2)
T_{H}(x_i, y_{i}, Q^2) T(x_{i}, \xi, t, Q^2)\;,
\label{amp}
\end{equation}
where $\phi(y_i,Q^2)$ is the antiproton distribution amplitude,
$T_{H}$ the hard scattering amplitude, calculated in the collinear approximation and $T(x_{i}, \xi, t, Q^2)$
the new TDAs.


%
%

To define the TDAs we introduce light-cone coordinates
$v^\pm = (v^0 \pm v^3) /\sqrt{2}$ and transverse components $v_T =
(v^1, v^2)$ for any four-vector $v$.  The skewness variable $\xi =
(p-p')^+ /(p+p')^+$ describes the loss of plus-momentum of the
incident nucleon and is connected with $x_B$  by $\xi \approx
{x_B}/{(2-x_B)}$ 

For instance, we define the leading twist TDAs for the $p \to \pi^0$ transition as :
\bea{TDA}
 &&  4  \langle     \pi^{0}(p')|\, 
\epsilon^{ijk}u^{i}_{\alpha}(z_1\,n) u^{j}_{\beta}(z_2\,n)d^{k}_{\gamma}(z_3\,n)
\,|p(p,s) \rangle 
\\ \nonumber
&&= -\frac{f_N}{2f_\pi}\left[ V^{0}_{1} (\hat P C)_{\alpha\beta}(B)_{\gamma}  +
A^{0}_{1} (\hat P\gamma^5 C)_{\alpha\beta}(\gamma^5 B)_{\gamma} -
3\,T^{0}_{1} ( P^\nu i\sigma_{\mu\nu} C)_{\alpha\beta}(\gamma^\mu B)_{\gamma} \right] 
\nonumber \\
&&+ V^{0}_{2} 
 (\hat P C)_{\alpha\beta}(\hat \Delta_{T} B)_{\gamma} +
A^{0}_{2}(\hat P \gamma^5 C)_{\alpha\beta}(\hat \Delta_{T}\gamma^5 B)_{\gamma}
+ T^{0}_{2} (\Delta_{T}^\mu P^\nu\, i\,\sigma_{\mu\nu} C)_{\alpha\beta}(B)_{\gamma}
\nonumber \\
&&+  T^{0}_{3} ( P^\nu \sigma_{\mu\nu} C)_{\alpha\beta}(\sigma^{\mu\rho}
\Delta_{T}^\rho B)_{\gamma} + \frac{T_{4}^0}{M } (\Delta_{T}^\mu P^\nu\, 
i\,\sigma_{\mu\nu} C)_{\alpha\beta}(\hat \Delta_{T}¥B)_{\gamma}\;, \nonumber
\eea
where $\sigma^{\mu\nu}= i/2[\gamma^\mu, \gamma^\nu]$, $C$ is the charge 
conjugation matrix 
and $B$ the nucleon spinor. 
$\hat P = P^\mu \gamma_\mu$, the vector $\Delta =p'-p$ has 
 - in the massless limit - the  transverse components 
$$
\Delta_T^\mu = (g^{\mu\nu} - \frac{1}{Pn}(P^\mu n^\nu +P^\nu n^\mu))\Delta_\nu\;.
$$
$f_\pi$ is the pion decay constant ( $f_\pi = 93$ MeV) and $f_N$ is the
constant which determines the value of the nucleon wave function at the
origin.
The first three 
terms in (\ref{TDA}) are the only ones surviving the forward limit
$\Delta_T \to 0$.
The constants in front of these three  
terms  have been chosen
in reference to  the soft pion limit results.
 With these conventions each function $V(z_{i}P\cdot n)$, $A(z_{i}P\cdot n)$,
$T(z_{i}P\cdot n)$ is dimensionless.

The TDAs can then be Fourier transformed to get the usual representation in terms of the 
momentum fractions, through the relation
\begin{equation}
F (z_{i}P\cdot n) = \int\limits^{1+\xi}_{-1+\xi} d^3x 
\delta (x_{1}+ x_{2}+ x_{3} -2\xi) e^{-iPn\Sigma x_{i}z_{i}} \, F(x_{i},\xi)
\end{equation}
where $F$ stands for $V_{i}, A_{i}, T_{i}$.

These TDAs are matrix elements of the same operator that appears in baryonic distribution amplitudes. They 
thus obey evolution equations which, as those of GPDs are of the ERBL type in some $x-$region, but are of a different nature in
another  $x-$region. 

At fixed $\xi$ and $t$ the scaling behaviour of the amplitude is easily derived from its factorized expression,
up to logarithmic corrections from the running of $\alpha_s$ and from the scale evolutions of the DA and of the TDA. 

A partonic understanding of backward VCS is thus available for the first time, and precise data may be collected 
at JLab soon. This will be more discussed in J.Ph. Lansberg's presentation \cite{LTrieste}.

\section{Impact picture}

\begin{figure}
\begin{center}
  \leavevmode
  \epsfxsize 0.6\textwidth
  \epsfbox{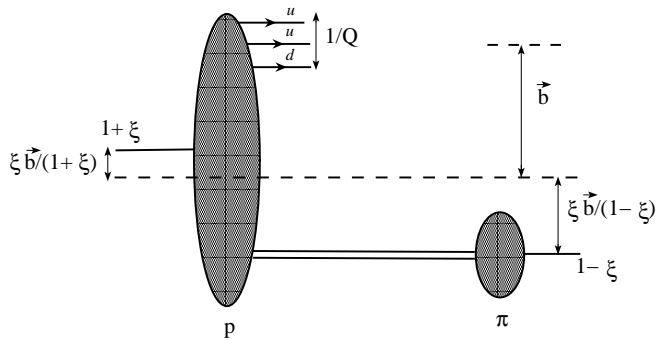}
\end{center}
\caption{\label{fig:TDA}The impact picture of the $p\to \pi $ TDA.}
\end{figure}

One of the particularly nice feature of this class of exclusive reactions is its ability to uncover
 the deep transverse structure of hadrons \cite{Impact}. GPDs, GDAs and TDAs contain information about the
spatial structure of hadrons.  For instance, the $p \to \pi $ TDA probes the partonic structure of the
proton by requiring its wave function to overlap with the wave
function of the configurations of the emerging meson, after it has been stripped from its valence quarks.
Moreover, the Fourier transform of its dependence
on $t$ tells us about the transverse position  of these valence quarks in the proton.
This may be phrased alternatively as detecting 
the transverse mean position of a pion inside the proton, 
when the proton 
state is of the "next to leading Fock " order, namely $ |\, qqq \,\pi >$. 
This is shown on Fig. \ref{fig:TDA}.

\section*{Acknowledgments}  
This work is supported by the Joint Research Activity "Generalised
Parton Distributions" in Integrated Infrastructure Initiative ``Hadron
Physics'' of the European Union, contract No.~RII3-CT-2004-506078 and by 
the Polish Grant 1 P03B 028 28.
L.Sz.\ is a Visiting Fellow of the Fonds National pour la Recherche
Scientifique (Belgium).  The work of B.P. and L. Sz.\ is partially
supported by the French-Polish scientific agreement Polonium.



\begin{thebibliography}{99}

\bibitem{Muller}
D.~M{\"u}ller {\it et al.}, 
Fortschr. Phys. {\bf 42}, 101 (1994); 
%
X.~D.~Ji,
Phys.\ Rev.\ Lett.\  {\bf 78}, 610 (1997); 
%
A.~V. Radyushkin,
Phys. Rev. {\bf D56}, 5524 (1997).

\bibitem{GDA}
  M.~Diehl {\it et al.},
Phys.\ Rev.\ Lett.\  {\bf 81}, 1782 (1998);
M.~Diehl, T.~Gousset and B.~Pire,
Phys.\ Rev.\ D {\bf 62}, 073014 (2000);
I.~V.~Anikin, B.~Pire and O.~V.~Teryaev,
Phys.\ Rev.\ D {\bf 69}, 014018 (2004) and  Phys.\ Lett.\ B {\bf 626}, 86 (2005).

\bibitem{TDA}
  B.~Pire and L.~Szymanowski,
  Phys.\ Rev.\ D {\bf 71} (2005) 111501;
  Phys.\ Lett.\ B {\bf 622} (2005) 83;
  J.~P.~Lansberg, B.~Pire and L.~Szymanowski,
  Phys.\ Rev.\ D {\bf 73} (2006) 074014.


\bibitem{asym}
  M.~Diehl {\it et al.}  Phys.\ Lett.\ B {\bf 411} (1997) 193;
A.~V.~Belitsky {\it et al.},
  Nucl.\ Phys.\ B {\bf 593} (2001) 289.
  
\bibitem{tGPD1} 
M.~Diehl, T.~Gousset and B.~Pire,
  Phys.\ Rev.\ D {\bf 59} (1999) 034023.


\bibitem{tGPD2} 
 D.~Y.~Ivanov {\it et al.},
  Phys.\ Lett.\ B {\bf 550} (2002) 65 and Phys.\ Part.\ Nucl.\  {\bf 35} (2004) S67.
 R.~Enberg, B.~Pire and L.~Szymanowski,
  arXiv:hep-ph/0601138.

\bibitem{Hybrid}
I.~V.~Anikin  {\it et al.},
  Phys.\ Rev.\ D {\bf 71} (2005) 034021,
  Phys.\ Rev.\ D {\bf 70} (2004) 011501 and  arXiv:hep-ph/0601176.

\bibitem{PAX}
  ``An  Int. Accelerator Facility for Beams  of Ions and
Antiprotons'', GSI Conceptual Design Report, Nov. 2001;
 V.~Barone {\it et al.}  [PAX Collaboration],
  arXiv:hep-ex/0505054.



\bibitem{Impact}
M.~Burkardt,
Phys.\ Rev.\ D {\bf 62}, 071503 (2000), 
Erratum-ibid.\ D {\bf 66}, 119903 (2002); 
J.~P.~Ralston and B.~Pire,
Phys.\ Rev.\ D {\bf 66}, 111501 (2002);
M.~Diehl,
Eur.\ Phys.\ J.\ C {\bf 25}, 223 (2002),
Erratum-ibid.\ C {\bf 31}, 277 (2003).
B.~Pire and L.~Szymanowski, Phys.\ Lett.\ B {\bf 556} (2003) 129 and PoS {\bf HEP2005}, 103 (2006).
\bibitem{LTrieste}  
 J.~P.~Lansberg, B.~Pire and L.~Szymanowski, these proceedings, 
arXiv: hep-ph/0607130.
\end{thebibliography}
\end{document}